\documentclass[11pt]{article}             
\usepackage{graphicx}
\newcommand{\bc}{\begin{center}}
\newcommand{\ec}{\end{center}}

\title{Qubit portrait of qudit states and Bell inequalities}
\author{V.~N.~Chernega$^{\dag}$, V.~I.~Man'ko$^{\ddag}$\\
\\$^{\dag}$Faculty of Physics, M.V. Lomonosov Moscow State University\\
Vorob'evy Gory, Moscow 119992, Russia
        \\$^{\ddag}$ P.N.~Lebedev Physical Institute, Russian Academy of Sciences\\
        Leninskii Prospect, 53, Moscow 119991, Russia \\
        \\Emails: vchernega@gmail.com, manko@sci.lebedev.ru}

\begin{document}

\maketitle

\begin{abstract}
A linear map of qudit tomogram onto qubit tomogram (qubit
portrait) is proposed as a characteristics of the qudit state.
Using the qubit portrait method the Bell inequalities for two
qubits and two qutrits are discussed in framework of probability
representation of quantum mechanics. Semigroup of stochastic
matrices is associated with tomographic probability distributions
of qubit and qutrit states. Bell-like inequalities are studied
using the semigroup of stochastic matrices. The qudit-qubit map of
tomographic probability distributions is discussed as ansatz to
provide a necessary condition for separability of quantum states.
\end{abstract}

\section{Introduction}

In probability representation of quantum states~\cite{Mancini96} the
states are described by probability distributions. For example, the
spin states are described by probability distribution ( called spin
tomogram)
$w(m,\overrightarrow{n})$~\cite{DodPhysLett}~\cite{OlgaJETP} where
$m$ is spin projection on direction determined by unit vector
$\overrightarrow{n}$. The role of spin tomograms for studying
separability and entanglement of quantum states was pointed out
in~\cite{SudPhysLet2004}.The aim of our work is to study properties
of spin tomograms for one and two spins. In quantum information
framework~\cite{Mendesquant-ph} we study qubits and qudits in the
context of separable and entangled states. We will obtain that the
separable two-qubit states can be associated with 4x4 - stochastic
matrices which form a semigroup. This property provides the Bell
inequality~\cite{Bell64},~\cite{CHSH} which serves as a criterion of
the separability. The Bell inequalities were considered in context
of the probability representation
in~\cite{Lupo},~\cite{Andreev2},~\cite{Andreev3}. The probability
representation for spin states was discussed and developped
in~\cite{Andreev1}, \cite{JRLR6}, \cite{JRLR7}, \cite{JRLR8},
\cite{JRLR13}, \cite{JRLR14}. The Shanon entropy~\cite{JRLR24} of
spin states was considered in~\cite{JRLR27}, \cite{JRLR30}. A linear
map of spin tomographic probability distribution (called
qudit-tomogram)onto qubit tomogram is constructed. The map provides
qubit portrait of qudit states. The qubit portrait is used to get
necessary condition of multiqudit state separability. The
preliminary remarks on such map were presented
in~\cite{JRLROlga2006}. We will discuss as examples some multiqudit
states. The paper is organized as follows: In Sec.2 we review
properties of stochastic matrices. In Sec.3 we derive an inequality
to be used for studying Bell inequality. In Sec.4 we consider
matrices as vectors. In Sec.5 we give a geometrical picture
associated with probabilities. In Sec.6 we give example of 3x3
stochastic matrices. In Sec.7 we present example of qubit states. In
Sec.8 we discuss entangled two-qubit states. In Sec.9 we formulate
new separability criterion related to semigroup of stochastic
matrices. In Sec.10 a new necessary condition of separability is
suggested. In Sec.11 example of two-qubit entangled state is
considered. In Sec.12 qubit portrait method is applied to
qubit-qutrit state. In Sec.13 concrete example is given. In Sec.14
general reduction criterion of separability is formulated. In Sec.15
conclusions and perspectives are discussed.

\section{Qubits and stochastic matrices}
For one qubit (or for the spin one-half particle state) any state
vector $|\psi\rangle$ has the form
\begin{equation}\label{eq.1}
|\psi\rangle=\pmatrix{a\cr b},\quad
\left(\langle\psi|=\pmatrix{a^*,b^*}\right),
\end{equation}
where the complex numbers $a=a_1+\iota a_2$ and $b=b_1+\iota b_2$
satisfy the normalization condition
\begin{equation}\label{eq.2}
\langle\psi|\psi\rangle=|a|^2+|b|^2=1.
\end{equation}
The 2x2-density matrix of the pure state $|\psi\rangle$ reads
\begin{equation}\label{eq.3}
\rho_{\psi}=|\psi\rangle\langle\psi|=\pmatrix{|a|^2&a b^* \cr b
a^*&|b|^2}.
\end{equation}
The trace of the density matrix is
\begin{equation}\label{eq.4}
\mbox{Tr}\rho_{\psi}=|a|^2+|b|^2=1.
\end{equation}
The diagonal elements of the density matrix determine the
probabilities for spin projections on z-axis $m=+1/2$ and
$m=-1/2$, i.e.
\begin{equation}\label{eq.5}
w(+\frac{1}{2})=|a|^2, \quad w(-\frac{1}{2})=|b|^2.
\end{equation}
Since the probabilities satisfy condition~(\ref{eq.2}) they can be
parameterized as follows
\begin{equation}\label{eq.6}
|a|^2=\cos^2\Theta, \quad |b|^2=\sin^2\Theta.
\end{equation}
Let us introduce the matrix
\begin{equation}\label{eq.7}
M=\left(%
\begin{array}{cc}
  p & q \\
  1-p & 1-q \\
\end{array}%
\right),
\end{equation}
where the real numbers $p$ and $q$ satisfy the inequalities
\begin{equation}\label{eq.8}
1\geq p\geq 0,\quad 1\geq q\geq 0.
\end{equation}
The nonnegative numbers $p$, 1-$p$ and $q$, 1-$q$ can be
considered as probability distributions. Numerical example of such
matrix reads
\begin{equation}\label{eq.9}
M_N=\left(%
\begin{array}{cc}
  \frac{1}{10} & \frac{2}{5} \\
  \frac{9}{10} & \frac{3}{5} \\
\end{array}%
\right).
\end{equation}
There are two probability distributions. Firsts one is
$(1/10,9/10)$. The second one is $(2/5,3/5)$. Important property
of the set of the matrices M is that the product of two matrices
of the form (\ref{eq.7}) has the same form, i.e.
\begin{equation}\label{eq.10}
M_1M_2=\left(%
\begin{array}{cc}
  p_1 & q_1 \\
  1-p_1 & 1-q_1 \\
\end{array}%
\right)~
\left(%
\begin{array}{cc}
  p_2 & q_2 \\
  1-p_2 & 1-q_2 \\
\end{array}%
\right)=\left(%
\begin{array}{cc}
  p_3 & q_3 \\
  1-p_3 & 1-q_3 \\
\end{array}%
\right)~.
\end{equation}
Here
\begin{eqnarray}
&&p_3=p_1p_2+q_1(1-p_2),\nonumber\\
&& q_3=p_1q_2+q_1(1-q_2).\label{eq.11}
\end{eqnarray}
The set of matrices (\ref{eq.7}) forms the semigroup. The unit
matrix belongs to the set. The inverse matrix
\begin{equation}\label{eq.12}
M^{-1}=\frac{1}{\mbox{det}M}\left(%
\begin{array}{cc}
  1-q & -q \\
  p-1 & p \\
\end{array}%
\right), \quad \mbox{det}M=p(1-q)-q(1-p),
\end{equation}
does not satisfy the condition (\ref{eq.8}) and does not belong to
the set of matrices (\ref{eq.7}). The subset of stochastic
matrices of the form
\begin{equation}\label{eq.13}
N=\left(%
\begin{array}{cc}
  p & 1-p \\
  1-p & p \\
\end{array}%
\right)
\end{equation}
is also the semigroup. In fact
\begin{equation}\label{eq.14}
N_1N_2=\left(%
\begin{array}{cc}
  p_1 & 1-p_1 \\
  1-p_1 & p_1 \\
\end{array}%
\right)~
\left(%
\begin{array}{cc}
  p_2 & 1-p_2 \\
  1-p_2 & p_2 \\
\end{array}%
\right)=\left(%
\begin{array}{cc}
  p_3 & 1-p_3 \\
  1-p_3 & p_3 \\
\end{array}%
\right)~,
\end{equation}
where the nonnegative number
\begin{equation}\label{eq.15}
p_3=p_1p_2+(1-p_1)(1-p_2)
\end{equation}
determines the matrix elements of the matrix
\begin{equation}\label{eq.16}
N_3=\left(%
\begin{array}{cc}
  p_3 & 1-p_3 \\
  1-p_3 & p_3 \\
\end{array}%
\right).
\end{equation}
The set of matrices (\ref{eq.13}) is called semigroup of
bistochastic matrices. The sum of numbers both in columns and in
rows of bistochastic matrices is equal to one. The bistochastic
matrices can be associated with $n$x$n$-unitary matrices $u$ with
matrix elements $u_{jk}$ satisfying the condition.
\begin{equation}
\sum_{k=1}^{n}\mid u_{jk}\mid^{2}=1, \quad \sum_{j=1}^{n}\mid
u_{jk}\mid^{2}=1.
\end{equation}
Thus the stochastic matrix $\wp$ with matrix elements.
\begin{equation}
\wp_{jk}=\mid u_{jk}\mid^{2}
\end{equation}
is the bistochastic matrix. It means that the group $u(n)$ of
unitary $n$x$n$ matrices induces the semigroup of bistochastic
matrices $\wp_{jk}=\mid u_{jk}\mid^{2}$. The tensor product of two
bistochastic matrices is a bistochastic matrix. Thus the group of
tensor product of unitary matrices $u(n_1)\bigotimes u(n_2)$
creates the semigroup which is tensor product of bistochastic
matrices $\wp_1\bigotimes\wp_2$ with matrix elements $\mid
u(n_1)_{jk}\mid^{2}$ and $\mid u(n_2)_{\alpha\beta}\mid^{2}$.
Using the property (\ref{eq.14}) one can introduce the associative
product of probability distributions. In fact given two
probability distributions $p_1,1-p_1$ and $p_2,1-p_2$. One can
associate with the probability distributions two vectors
\begin{equation}
\overrightarrow{w_1}=\left(%
\begin{array}{c}
  p_1 \\
  1-p_1 \\
\end{array}%
\right)\equiv \left(%
\begin{array}{c}
  w_1^{(1)} \\
  w_2^{(1)} \\
\end{array}%
\right),
\end{equation}

\begin{equation}
\overrightarrow{w_2}=\left(%
\begin{array}{c}
  p_2 \\
  1-p_2 \\
\end{array}%
\right)\equiv \left(%
\begin{array}{c}
  w_1^{(2)} \\
  w_2^{(2)} \\
\end{array}%
\right),
\end{equation}

and two matrices

\begin{equation}
N_1=\left(%
\begin{array}{cc}
  w_1^{(1)} & w_2^{(1)} \\
  w_2^{(1)} & w_1^{(1)} \\
\end{array}%
\right),
\end{equation}

\begin{equation}
N_2=\left(%
\begin{array}{cc}
  w_1^{(2)} & w_2^{(2)} \\
  w_2^{(2)} & w_1^{(2)} \\
\end{array}%
\right).
\end{equation} We define the associative product
$\overrightarrow{w_3}$ of two vectors (called star-product)
$\overrightarrow{w_1}*\overrightarrow{w_2}=\overrightarrow{w_3}$
using the result of multiplication of two matrices $N_1$ and $N_2$
given by (\ref{eq.14}) and (\ref{eq.15}) for finding the component
of the vector $\overrightarrow{w_3}$. We get
\begin{equation}
w_1^{(3)}=w_1^{(1)}w_1^{(2)}+w_2^{(1)}w_2^{(2)},
\end{equation}
\begin{equation}
w_2^{(3)}=w_2^{(1)}w_1^{(2)}+w_1^{(1)}w_2^{(2)}.
\end{equation}
This result can be generalized to introduce the associative
product by means of the same tools for N-dimensional vectors. The
components of the product vector read
\begin{equation}
p_m=\sum_{k=1}^{N} w_{[k+m-1]_{N}}W_k.
\end{equation}
Here $[k+m-1]_{N}$ means the number
\begin{equation}
\left\{%
\begin{array}{ll}
    k+m-1, & \hbox{if} \quad k+m-1 < N; \\
    k+m-1-N, & \hbox{if} \quad k+m-1 > N. \\
\end{array}%
\right.
\end{equation}

The eigenvalues of the stochastic matrix (\ref{eq.7}) are
\begin{eqnarray}\label{eq.17}
\lambda_1=1,\quad \lambda_2=p-q.
\end{eqnarray}
They satisfy the condition
\begin{equation}\label{eq.18}
|\lambda_k|\leq 1,\quad k=1,2.
\end{equation}
The eigenvectors of the matrix (\ref{eq.7}) read
\begin{eqnarray}
&&|U_1\rangle=\left(%
\begin{array}{c}
  1 \\
  q^{-1}(1-p) \\
\end{array}%
\right);\nonumber\\
&&|U_{p-q}\rangle=\left(%
\begin{array}{c}
  -1 \\
  1 \\
\end{array}%
\right). \label{eq.19}
\end{eqnarray}
It means that the matrix $M$ can be presented in the form
\begin{equation}\label{eq.20}
\left(%
\begin{array}{cc}
  p & q \\
  1-p & 1-q \\
\end{array}%
\right)=U\left(%
\begin{array}{cc}
  1 & 0 \\
  0 & p-q \\
\end{array}%
\right)U^{-1},
\end{equation}
where the matrix $U$ reads
\begin{equation}\label{eq.21}
U=\left(%
\begin{array}{cc}
  1 & 1 \\
  q^{-1}(1-p) & -1 \\
\end{array}%
\right).
\end{equation}
In the case $p=q$ the determinant of the stochastic matrix equals
zero. The inverse matrix has the form
\begin{equation}\label{eq.22}
U^{-1}=\frac{1}{1+q^{-1}(1-p)}\left(%
\begin{array}{cc}
  1 & 1 \\
  q^{-1}(1-p) & -1 \\
\end{array}%
\right).
\end{equation}
It means that
\begin{equation}\label{eq.23}
U^{-1}=U\frac{1}{1+q^{-1}(1-p)}\,,
\end{equation}
and
\begin{equation}\label{eq.24}
U^2=(1+q^{-1}(1-p))\left(%
\begin{array}{cc}
  1 & 0 \\
  0 & 1 \\
\end{array}%
\right).
\end{equation}
From this property follows
\begin{equation}\label{eq.25}
U^{2k}=(1+q^{-1}(1-p))^k\left(%
\begin{array}{cc}
  1 & 0 \\
  0 & 1 \\
\end{array}%
\right)\,,
\end{equation}
\begin{equation}\label{eq.26}
U^{2k+1}=(1+q^{-1}(1-p))^kU.
\end{equation}
From  (\ref{eq.20}) we get
\begin{equation}\label{eq.28}
\left(%
\begin{array}{cc}
  p & q \\
  1-p & 1-q \\
\end{array}%
\right)^n=U\left(%
\begin{array}{cc}
  1 & 0 \\
  0 & (p-q)^n \\
\end{array}%
\right)U^{-1}\,, \quad n=1,2,3...
\end{equation}
Since $|p-q|\leq1$, for large $n$ $|(p-q)|^n\ll1$ In this case the
matrix
\begin{equation}\label{eq.29}
\left(%
\begin{array}{cc}
  1 & 0 \\
  0 & (p-q)^n \\
\end{array}%
\right)\rightarrow \left(%
\begin{array}{cc}
  1 & 0 \\
  0 & 0 \\
\end{array}%
\right).
\end{equation}

\section{Useful inequality}
We prove now an useful inequality for scalar product of two pairs
of real vectors. Let

\begin{eqnarray}\label{eq.30}
|(\overrightarrow{a_1}\overrightarrow{b_1})|<c\quad \mbox{and}
\quad |(\overrightarrow{a_2}\overrightarrow{b_2})|<c,
\end{eqnarray}
where $c$ is a positive number. Then the convex sum
$\cos^2\gamma(\overrightarrow{a_1}\overrightarrow{b_1})+\sin^2\gamma(\overrightarrow{a_2}\overrightarrow{b_2})$
satisfies the inequality
\begin{eqnarray}\label{eq.31}
|\cos^2\gamma(\overrightarrow{a_1}\overrightarrow{b_1})+\sin^2\gamma(\overrightarrow{a_2}\overrightarrow{b_2})|<c.
\end{eqnarray}
By induction we get the inequality for generic convex sum. If
$|\overrightarrow{a_k}\overrightarrow{b_k}|<c$, then
\begin{eqnarray}\label{eq.32}
|\sum_k p_k(\overrightarrow{a_k}\overrightarrow{b_k})|<c,
\end{eqnarray}
where the coefficients
\begin{eqnarray}\label{eq.33}
1\geq p_k\geq0, \quad\sum_k p_k=1.
\end{eqnarray}
In particular, we get the following inequality. If
\overrightarrow{b_1}=\overrightarrow{b_2}=...=\overrightarrow{b_k}=...=\overrightarrow{B}
the property (\ref{eq.32}) reads
\begin{eqnarray}\label{eq.34}
|\sum_k p_k(\overrightarrow{a_k}\overrightarrow{B})|<c,
\end{eqnarray}
i.e
\begin{eqnarray}\label{eq.35}
|\sum_k((p_k\overrightarrow{a_k})\overrightarrow{B})|<c.
\end{eqnarray}

\section{Matrices as vectors}
We discuss below how matrices can be interpreted as vectors. For
example, the real 2x2 matrix
\begin{eqnarray}\label{eq.36}
\mu=\left(%
\begin{array}{cc}
  a & b \\
  c & d \\
\end{array}%
\right)
\end{eqnarray}
can be considered as the vector
\begin{eqnarray}\label{eq.37}
\overrightarrow{\mu}=\left(%
\begin{array}{c}
  a \\
  b \\
  c \\
  d \\
\end{array}%
\right).
\end{eqnarray}
The sum of two matrices $\mu_1$ and $\mu_2$
\begin{eqnarray}\label{eq.38}
\mu_1+\mu_2=\left(%
\begin{array}{cc}
  a_1+a_2 & b_1+b_2 \\
  c_1+c_2 & d_1+d_2 \\
\end{array}%
\right)
\end{eqnarray}
can be interpreted as sum of two vectors with following components
\begin{eqnarray}\label{eq.39}
\overrightarrow{\mu_1}+\overrightarrow{\mu_2}=\left(%
\begin{array}{c}
  a_1+a_2 \\
  b_1+b_2 \\
  c_1+c_2 \\
  d_1+d_2 \\
\end{array}%
\right).
\end{eqnarray}
Then the number $Tr(\mu_1^{tr}\mu_2)=a_1a_2+b_1b_2+c_1c_2+d_1d_2$
where $\mu_1^{tr}$ is transposed matrix $\mu_1$, is standard
scalar product of two vectors, i.e.
\begin{eqnarray}\label{eq.40}
\mbox{Tr}(\mu_1^{tr}\mu_2)=(\overrightarrow{\mu_1}\overrightarrow{\mu_2}).
\end{eqnarray}
Let us make a remark. The stochastic matrix $M$ (\ref{eq.7})
becomes new stochastic matrix $M'$ if one permutes the columns of
the matrix $M$, i.e.
\begin{eqnarray}\label{eq.41}
M'=\left(%
\begin{array}{cc}
  q & p \\
  1-q & 1-p \\
\end{array}%
\right).
\end{eqnarray}
The same property takes place if one permutes rows of the matrix
$M$. In this case we get new stochastic matrix
\begin{eqnarray}\label{eq.42}
M''=\left(%
\begin{array}{cc}
  1-p & 1-q \\
  p & q \\
\end{array}%
\right).
\end{eqnarray}
\section{Geometrical picture}
The probabilities $1\geq w_1\geq 0$ and $1\geq w_2\geq 0$ such that
$w_1 + w_2 = 1$ can be considered in geometrical terms as points on
a simplex which is the line shown in Fig.~1


As example we show vector with its end posed on the line and it can
be given as the column
\begin{eqnarray}
\overrightarrow{w}=\left(%
\begin{array}{c}
  w_1 \\
  w_2 \\
\end{array}%
\right).
\end{eqnarray}
The stochastic matrices transform the vector $\overrightarrow{w}$
into another vector $\overrightarrow{W}$, for example
\begin{eqnarray}
\overrightarrow{W}=M\overrightarrow{w}.
\end{eqnarray}
One can check that the components of vector
\begin{eqnarray}
\left(%
\begin{array}{c}
  W_1 \\
  W_1 \\
\end{array}%
\right)=\left(%
\begin{array}{cc}
  q & p \\
  1-q & 1-p \\
\end{array}%
\right)\left(%
\begin{array}{c}
  w_1 \\
  w_2 \\
\end{array}%
\right)
\end{eqnarray}
satisfy the conditions $1\geq W_1 \geq 0$,\quad $1\geq W_2 \geq 0$,
\quad $W_1+W_2=1$.It means that the stochastic matrices move the
initial point on the simplex into another point on the same simplex.
The new probability distribution described by the vector
$\overrightarrow{W}$ has the components $W_1=qw_1+pw_2$,\quad
$W_2=(1-q)w_1+(1-p)w_2$. For bistochastic matrices, one has the
transformation $W_1=qw_1+(1-q)w_2$ $\quad W_2=(1-q)w_1+qw_2$. The
point $w_1=1/2$, $\quad w_2=1/2$ is invariant under this action. For
distributions with three components, the simplex has the geometrical
form of the plane shown in Fig.~2.


All the points on the triangle shown on this figure correspond to
all the probability distributions with three outputs. Below we
discuss the stochastic matrices which transform point on this
simplex into another point of the same simplex.

\section{The 3x3-stochastic matrices and linear maps of distributions}
Let us discuss now the stochastic matrices of the 3rd order of the
form
\begin{eqnarray}\label{eq.43}
M=\left(%
\begin{array}{ccc}
  p_1 & q_1 & r_1 \\
  p_2 & q_2 & r_2 \\
  p_3 & q_3 & r_3 \\
\end{array}%
\right).
\end{eqnarray}
Here the positive numbers $p_k,q_k,r_k\quad(k=1,2,3)$ satisfy the
normalization conditions
\begin{eqnarray}\label{eq.44}
\sum_{k=1}^{3}p_k= \sum_{k=1}^{3}q_k= \sum_{k=1}^{3}r_k=1.
\end{eqnarray}
It means that the numbers in columns of the matrix $M$ can be
interpreted as probability distributions. It is easy to check that
the set of all the matrices $M$ (\ref{eq.43}) form semigroup. Let
us give numerical example of such a matrix, i.e.
\begin{eqnarray}\label{eq.45}
M=\left(%
\begin{array}{ccc}
  \frac{1}{10} & \frac{1}{3} & \frac{8}{10} \\
  \frac{3}{10} & 0 & \frac{1}{10} \\
  \frac{6}{10} & \frac{2}{3} & \frac{1}{10} \\
\end{array}%
\right).
\end{eqnarray}
It is interesting that the eigenvalues of the stochastic matrix
$M$ contain $\lambda_1=1$. This eigenvalue 1 have stochastic
matrices $M_N$ of all dimensions $N\geq2$. One can see that other
eigenvalues of the stochastic matrix $M_N$ can be either real or
complex. Also all the eigenvalues of the stochastic matrices $M_N$
satisfy inequality $|\lambda_k|\leq1,\quad k=1,2,...,N$. We point
out that the permutations of elements of a chosen column transform
the stochastic matrix into another stochastic matrix. The group of
all permutations of matrix elements of $M_N$ - stochastic matrix
has $(N!)^{N+1}$ symmetry elements. The group elements are
independent permutations in each column $(N!)^N$ combined with
$N!$ permutation of columns . Trace of stochastic matrix $M_N$
satisfies inequality $\mbox{Tr}M_N\leq N$. The bistochastic 3x3 -
matrices have the form (\ref{eq.43}) but satisfy extra condition
$p_k+q_k+r_k =1 \quad(k=1,2,3)$. The discussed stochastic and
bistochastic matrices move the points on the triangle. The point
with components $(1/3,1/3,1/3)$ is invariant under the action of
the bistochastic matrices. Let us consider the first column of
3x3-stochastic matrix (\ref{eq.43}). The nonnegative matrix
elements in this column $p_1,p_2,p_3$ can be mapped onto three
pairs of nonnegative numbers:
\begin{eqnarray}\label{eq.aG1}
P_{1}^{(1)}=p_1, \quad P_{2}^{(1)}=(p_2+p_3);
\end{eqnarray}
\begin{eqnarray}\label{eq.aG2}
P_{1}^{(2)}=p_1+p_2, \quad P_{2}^{(2)}=p_3;
\end{eqnarray}
\begin{eqnarray}\label{eq.aG3}
P_{1}^{(3)}=p_1+p_3,\quad P_{2}^{(3)}=p_2.
\end{eqnarray}
Thus we get three probability distributions
$(P_{1}^{(1)},P_{2}^{(1)})$; $(P_{1}^{(2)},P_{2}^{(2)})$;
$(P_{1}^{(3)},P_{2}^{(3)})$ and distributions obtained by
permutations of these numbers. One can see that we constructed the
linear map of initial probability distribution with three possible
outcomes onto a set of probability distributions with two
outcomes. The map is invertible. In fact
\begin{eqnarray}\label{eq.aG4}
p_1=P_{1}^{(1)}, \quad p_2=P_{1}^{(2)}-P_{1}^{(1)}, \quad
p_3=P_{2}^{(2)}.
\end{eqnarray}
It means that knowing two probability distributions (\ref{eq.aG1})
-(\ref{eq.aG2}) we can reconstruct the initial distribution. We
call the set of probability distributions (\ref{eq.aG1}) and
(\ref{eq.aG2}) as qubit "portrait" of initial qutrit distribution.
We introduce this terminology because we will apply the
constructed map to study necessary conditions of separability for
quantum multiqudit states. Using the suggested ansatz one can
construct the analogous map for obtaining analogous portraits of
joint probability distributions.

\section{Qubit}
If one takes a convex sum of pure state density matrices we get
the density matrix of mixed state of spin $-1/2$ particle (or
qubit state). It means that the matrix
\begin{eqnarray}\label{eq.46}
\rho=\sum_k p_k|\psi_k\rangle\langle\psi_k|,
\end{eqnarray}
where $1\geq p_k \geq 0$ and $\sum_{k} p_k=1$ is hermitian matrix
\begin{eqnarray}\label{eq.47}
\rho^+=\rho,
\end{eqnarray}
and its trace is equal to 1. The density matrix is nonnegative
matrix, i.e. its eigenvalues are nonnegative numbers. The tomogram
of the qubit state is defined by formula
\begin{eqnarray}\label{eq.48}
w(m,U)=\left(%
\begin{array}{c}
  w(+\frac{1}{2},U) \\
  w(-\frac{1}{2},U) \\
\end{array}%
\right)=(U^+\rho U)_{mm}.
\end{eqnarray}
Here $U$ is unitary matrix. It has the form
\begin{eqnarray}\label{eq.49}
U=\left(%
\begin{array}{cc}
  \cos\frac{\theta}{2} e^{\frac{\iota(\varphi+\psi)}{2}} & \sin\frac{\theta}{2} e^{\frac{\iota(\varphi-\psi)}{2}} \\
  -\sin\frac{\theta}{2} e^{\frac{-\iota(\varphi-\psi)}{2}} & \cos\frac{\theta}{2} e^{\frac{-\iota(\varphi+\psi)}{2}} \\
\end{array}%
\right),
\end{eqnarray} and $\varphi,\theta,\psi$ are Euler angles.
In reality the Euler angle $\psi$ is not present in the final
expression of the tomogram. The tomogram is probability
distribution. In our previous notations we can introduce
stochastic matrix using substitutions
\begin{eqnarray}
&&p=w(+\frac{1}{2},U_1),
\nonumber\\
&&q=w(+\frac{1}{2},U_2), \label{eq.50}
\end{eqnarray}
i.e.
\begin{eqnarray}\label{eq.51}
M=\left(%
\begin{array}{cc}
  w(+\frac{1}{2},U_1) & w(+\frac{1}{2},U_2) \\
  1-w(+\frac{1}{2},U_1) & 1-w(+\frac{1}{2},U_2) \\
\end{array}%
\right).
\end{eqnarray}
Here $U_1$ is matrix determined by angels
$\varphi_1,\theta_1,\psi_1$ and the matrix $U_2$ is determined by
angels $\varphi_2,\theta_2,\psi_2$. The constructed stochastic
matrix with matrix elements equal to tomographic probabilities has
all the properties of stochastic matrices (\ref{eq.7}) discussed
in previous sections.

\section{Two qubits, separable and entangled states}
Let us introduce a unit vector
$\overrightarrow{n}=(\sin\theta\cos\varphi, \sin\theta\sin\varphi,
\cos\theta)$ which is normal vector to sphere surface. The
tomogram $w(m,U)$ can be considered as function on the sphere
\begin{eqnarray}\label{eq.52}
w(m,U)\equiv w(m,\overrightarrow{n}).
\end{eqnarray}
The stochastic matrix $M$ can be rewritten in the form
\begin{eqnarray}\label{eq.53}
M=\left(%
\begin{array}{cc}
  w(+\frac{1}{2},\overrightarrow{n_1}) & w(+\frac{1}{2},\overrightarrow{n_2}) \\
  1-w(+\frac{1}{2},\overrightarrow{n_1}) & 1-w(+\frac{1}{2},\overrightarrow{n_2}) \\
\end{array}%
\right).
\end{eqnarray}
One has
\begin{eqnarray}
&&w(-\frac{1}{2},\overrightarrow{n_1})=1-w(+\frac{1}{2},\overrightarrow{n_1}),
\nonumber\\[-2mm]
&&\label{eq.54}\\[-2mm]
&&w(-\frac{1}{2},\overrightarrow{n_2})=1-w(+\frac{1}{2},\overrightarrow{n_2}).
\nonumber
\end{eqnarray}
Let us consider two qubits. It means that we consider 4x4-density
matrix $\rho$. The tomogram of two-qubit state reads
\begin{eqnarray}\label{eq.55}
w(m_1,m_2,\overrightarrow{n},\overrightarrow{N})=(U^{\dag}\rho
U)_{m_1m_2,m_1m_2}.
\end{eqnarray}
Here $U$ is 4x4 unitary matrix which is tensor product of two 2x2
- unitary matrices
\begin{eqnarray}\label{eq.56}
U=U_1\bigotimes U_2,
\end{eqnarray}
where $U_1$ and $U_2$ are given by formula (\ref{eq.49}) with
Euler angels $\varphi_1\theta_1\psi_1$, $\varphi_2\theta_2\psi_2$,
respectively. The vector $\overrightarrow{n}$ is determined be
Euler angels $\varphi_1\theta_1\psi_1$ and vector
$\overrightarrow{N}$ is determined by Euler angels
$\varphi_2\theta_2\psi_2$. Simply separable state has the tomogram
of the factorized form
$w(m_1,m_2,\overrightarrow{n},\overrightarrow{N})=w_1(m_1,\overrightarrow{n})w_2(m_2,\overrightarrow{N})$.
Let us construct 4x4- stochastic matrix by following rule. We take
4 vectors
$\overrightarrow{a},\overrightarrow{b},\overrightarrow{c},\overrightarrow{d}$.
Then we choose 2 vectors $\overrightarrow{n}$ to be equal
$\overrightarrow{a}$ and $\overrightarrow{b}$ and 2 vectors
$\overrightarrow{N}$ to be equal $\overrightarrow{c}$ and
$\overrightarrow{d}$. We have two probability distributions for
first qubit $w_1(m_1,\overrightarrow{a})$,
$w_1(m_1,\overrightarrow{b})$ and two probability distributions
for second qubit $w_2(m_2,\overrightarrow{c})$,
$w_2(m_2,\overrightarrow{d})$. Then our 4x4-stochastic matrix
reads
\begin{eqnarray}\label{eq.57}
(M_4)_{k1}=\left(%
\begin{array}{c}
  w_1(+\frac{1}{2},\overrightarrow{a})w_2(+\frac{1}{2},\overrightarrow{b}) \\
  w_1(+\frac{1}{2},\overrightarrow{a})w_2(-\frac{1}{2},\overrightarrow{b}) \\
  w_1(-\frac{1}{2},\overrightarrow{a})w_2(+\frac{1}{2},\overrightarrow{b}) \\
  w_1(-\frac{1}{2},\overrightarrow{a})w_2(-\frac{1}{2},\overrightarrow{b}) \\
\end{array}%
\right),\quad k=1,2,3,4
\end{eqnarray}
\begin{eqnarray}\label{eq.58}
(M_4)_{k2}=\left(%
\begin{array}{c}
  w_1(+\frac{1}{2},\overrightarrow{a})w_2(+\frac{1}{2},\overrightarrow{c}) \\
  w_1(+\frac{1}{2},\overrightarrow{a})w_2(-\frac{1}{2},\overrightarrow{c}) \\
  w_1(-\frac{1}{2},\overrightarrow{a})w_2(+\frac{1}{2},\overrightarrow{c}) \\
  w_1(-\frac{1}{2},\overrightarrow{a})w_2(-\frac{1}{2},\overrightarrow{c}) \\
\end{array}%
\right),\quad k=1,2,3,4
\end{eqnarray}
\begin{eqnarray}\label{eq.59}
(M_4)_{k3}=\left(%
\begin{array}{c}
  w_1(+\frac{1}{2},\overrightarrow{d})w_2(+\frac{1}{2},\overrightarrow{b}) \\
  w_1(+\frac{1}{2},\overrightarrow{d})w_2(-\frac{1}{2},\overrightarrow{b}) \\
  w_1(-\frac{1}{2},\overrightarrow{d})w_2(+\frac{1}{2},\overrightarrow{b}) \\
  w_1(-\frac{1}{2},\overrightarrow{d})w_2(-\frac{1}{2},\overrightarrow{b}) \\
\end{array}%
\right),\quad k=1,2,3,4
\end{eqnarray}
and
\begin{eqnarray}\label{eq.60}
(M_4)_{k4}=\left(%
\begin{array}{c}
  w_1(+\frac{1}{2},\overrightarrow{d})w_2(+\frac{1}{2},\overrightarrow{c}) \\
  w_1(+\frac{1}{2},\overrightarrow{d})w_2(-\frac{1}{2},\overrightarrow{c}) \\
  w_1(-\frac{1}{2},\overrightarrow{d})w_2(+\frac{1}{2},\overrightarrow{c}) \\
  w_1(-\frac{1}{2},\overrightarrow{d})w_2(-\frac{1}{2},\overrightarrow{c}) \\
\end{array}%
\right),\quad k=1,2,3,4.
\end{eqnarray}
This matrix can be presented in the form of tensor product of two
stochastic 2x2-matrices, i.e.
\begin{eqnarray}\label{eq.61}
M_4=\left(%
\begin{array}{c}
  w_1(+\frac{1}{2},\overrightarrow{a})w_1(+\frac{1}{2},\overrightarrow{d}) \\
  w_1(-\frac{1}{2},\overrightarrow{a})w_1(-\frac{1}{2},\overrightarrow{d}) \\
\end{array}%
\right)\bigotimes
\left(%
\begin{array}{c}
  w_2(+\frac{1}{2},\overrightarrow{b})w_2(+\frac{1}{2},\overrightarrow{c}) \\
  w_2(-\frac{1}{2},\overrightarrow{b})w_2(-\frac{1}{2},\overrightarrow{c}) \\
\end{array}%
\right).
\end{eqnarray}
We call this stochastic matrix as "simply separable stochastic
matrix". One can check that the matrix (\ref{eq.61}) satisfies the
inequality ((Bell-CHSH) inequality~\cite{CHSH})

\begin{eqnarray}
&&|(M_4)_{11}-(M_4)_{21}-(M_4)_{31}+(M_4)_{41}+(M_4)_{12}-(M_4)_{22}-(M_4)_{32}+(M_4)_{42}\nonumber \\
&&+(M_4)_{13}-(M_4)_{23}-(M_4)_{33}+(M_4)_{43}-(M_4)_{14}+(M_4)_{24}+(M_4)_{34}-(M_4)_{44}|\leq
2.\label{eq.62}
\end{eqnarray}
This inequality can be rewritten in matrix form as
$|\mbox{Tr}(M_4I)|\leq 2$ \\where

\begin{eqnarray}\label{eq.63}
I=\left(%
\begin{array}{cccc}
  1 & -1 & -1 & 1 \\
  1 & -1 & -1 & 1 \\
  1 & -1 & -1 & 1 \\
  -1 & 1 & 1 & -1 \\
\end{array}%
\right).
\end{eqnarray}
The inequality has to be preserved if one changes the matrix $I$
by the product matrix $\tilde{I}=IC$, $C=C_1\bigotimes C_2$. Here
two 2x2-matrices $C_1$ and $C_2$ are arbitrary stochastic
matrices. In vector form $M_4\rightarrow \overrightarrow{M_4}$ and
according to rules of Sec.3 $I\rightarrow \overrightarrow{I}$ this
inequality reads
\begin{eqnarray}\label{eq.64}
|(\overrightarrow{I}\overrightarrow{M_4})|\leq 2.
\end{eqnarray}
Due to property of convex sums (\ref{eq.32}) one can state that if
one constructs a convex sum of matrices of the type $M_4$
\begin{eqnarray}\label{eq.65}
M=\sum_kP_kM_{4}^{(k)}, \quad P_k\geq0, \quad \sum_kP_k=1;
\end{eqnarray}
we get inequality
\begin{eqnarray}\label{eq.66}
|\overrightarrow{I}\overrightarrow{M}|\leq 2;
\end{eqnarray}
or
\begin{eqnarray}\label{eq.67}
|\mbox{Tr}(MI)|\leq 2.
\end{eqnarray}

\section{Separable and entangled states}
By definition the quantum state of two qubits is separable if the
tomogram of the state can be presented in the form of convex sum
of simply separable tomograms, i.e.
\begin{eqnarray}\label{eq.68}
w(m_1m_1\overrightarrow{n_1}\overrightarrow{n_2})=\sum_kP_kw_1^{(k)}(m_1\overrightarrow{n_1})w_2^{(k)}(m_2\overrightarrow{n_2});
\quad P_k\geq0; \quad \sum_kP_k=1.
\end{eqnarray}
Here the index $k$ can be understood as a collective index with
any number of components including both discrete and continuous
ones. One can see that the stochastic matrix corresponding to the
tomogram (\ref{eq.68}) has the form of convex sum of the matrices
of type (\ref{eq.61}), i.e.
\begin{eqnarray}\label{eq.69}
M_4=\sum_kP_k\left(%
\begin{array}{c}
  w^{(k)}(+\frac{1}{2},\overrightarrow{a})w^{(k)}(+\frac{1}{2},\overrightarrow{d}) \\
  w^{(k)}(-\frac{1}{2},\overrightarrow{a})w^{(k)}(-\frac{1}{2},\overrightarrow{d}) \\
\end{array}%
\right)\bigotimes
\left(%
\begin{array}{c}
  w^{(k)}(+\frac{1}{2},\overrightarrow{b})w^{(k)}(+\frac{1}{2},\overrightarrow{c}) \\
  w^{(k)}(-\frac{1}{2},\overrightarrow{b})w^{(k)}(-\frac{1}{2},\overrightarrow{c}) \\
\end{array}%
\right).
\end{eqnarray}
We call this stochastic matrix as "separable stochastic matrix".\\
$\underline{Lemma}$
\\The product of two stochastic matrices $M_4^{(1)},M_4^{(2)}$
corresponding to tomograms of separable states  of two qubits is
the convex sum of simply separable stochastic matrices.
\\$\underline{Proof}$\\
Let $F_1$ be stochastic matrix corresponding to separable two
qubit quantum state, i.e. it can be written in the form
(\ref{eq.69}) which we denote as
\begin{eqnarray}\label{eq.70}
F_1=\sum_kP_kw^{(k)}_{(1)}.
\end{eqnarray}
Here
\begin{eqnarray}\label{eq.71}
w^{(k)}_{(1)}=\left(%
\begin{array}{c}
  w^{(k)}(+\frac{1}{2},\overrightarrow{a_1})w^{(k)}(+\frac{1}{2},\overrightarrow{d_1}) \\
  w^{(k)}(-\frac{1}{2},\overrightarrow{a_1})w^{(k)}(-\frac{1}{2},\overrightarrow{d_1}) \\
\end{array}%
\right)\bigotimes
\left(%
\begin{array}{c}
  w^{(k)}(+\frac{1}{2},\overrightarrow{b_1})w^{(k)}(+\frac{1}{2},\overrightarrow{c_1}) \\
  w^{(k)}(-\frac{1}{2},\overrightarrow{b_1})w^{(k)}(-\frac{1}{2},\overrightarrow{c_1}) \\
\end{array}%
\right).
\end{eqnarray}
Let $F_2$ be another stochastic matrix of the form
\begin{eqnarray}\label{eq.72}
F_2=\sum_s\rho_sw_{(2)}^{(s)}.
\end{eqnarray}
Here $\rho_s\geq0, \sum_s\rho_s=1$ and notation (\ref{eq.72})
means that we change in (\ref{eq.71}) $k\rightarrow s,
\overrightarrow{a_1}\rightarrow
\overrightarrow{a_2},\overrightarrow{d_1}\rightarrow
\overrightarrow{d_2},\overrightarrow{b_1}\rightarrow
\overrightarrow{b_2},\overrightarrow{c_1}\rightarrow
\overrightarrow{c_2}$. Let us calculate the product matrix
\begin{eqnarray}\label{eq.73}
F=F_1F_2=\sum_{ks}(P_k\rho_s)w_{(1)}^{(k)}w_{(2)}^{(s)}.
\end{eqnarray}
Since the rule of multiplication of tensor products of matrices
reads
\begin{eqnarray}\label{eq.74}
(a\bigotimes b)(c\bigotimes d)=(ac)\bigotimes(bd),
\end{eqnarray}
one has
\begin{eqnarray}\label{eq.75}
F=\sum_jQ_jw^{j}.
\end{eqnarray}
Here $j$ is collective index $j=(ks)$, the matrix $w^{(j)}$ is the
4x4-stochastic matrix of simply separable form. It means that the
matrix $F$ satisfies the Bell-CHSH inequality
\begin{eqnarray}\label{eq.76}
|\mbox{Tr}(FI)|\leq2.
\end{eqnarray}

\section{Necessary condition of separability}
We will use this lemma to formulate the necessary condition of the
separability of two qubit state.
\\In fact if one has the
two qubit separable state with spin tomogram
$w(m_1m_2\overrightarrow{n_1}\overrightarrow{n_2})$ the set of
matrices associated with the tomogram using the following rule
\begin{eqnarray}\label{eq.77}
M(\overrightarrow{a}\overrightarrow{b}\overrightarrow{c}\overrightarrow{d})=\left(%
\begin{array}{cccc}
  w(+\frac{1}{2}\overrightarrow{a},+\frac{1}{2}\overrightarrow{b}) & w(+\frac{1}{2}\overrightarrow{a},+\frac{1}{2}\overrightarrow{c}) & w(+\frac{1}{2}\overrightarrow{d},+\frac{1}{2}\overrightarrow{b}) & w(+\frac{1}{2}\overrightarrow{d},+\frac{1}{2}\overrightarrow{c})\\
  w(+\frac{1}{2}\overrightarrow{a},-\frac{1}{2}\overrightarrow{b}) & w(+\frac{1}{2}\overrightarrow{a},-\frac{1}{2}\overrightarrow{c}) & w(+\frac{1}{2}\overrightarrow{d},-\frac{1}{2}\overrightarrow{b}) & w(+\frac{1}{2}\overrightarrow{d},-\frac{1}{2}\overrightarrow{c})\\
  w(-\frac{1}{2}\overrightarrow{a},+\frac{1}{2}\overrightarrow{b}) & w(-\frac{1}{2}\overrightarrow{a},+\frac{1}{2}\overrightarrow{c}) & w(-\frac{1}{2}\overrightarrow{d},+\frac{1}{2}\overrightarrow{b}) & w(-\frac{1}{2}\overrightarrow{d},+\frac{1}{2}\overrightarrow{c})\\
  w(-\frac{1}{2}\overrightarrow{a},-\frac{1}{2}\overrightarrow{b}) & w(-\frac{1}{2}\overrightarrow{a},-\frac{1}{2}\overrightarrow{c}) & w(-\frac{1}{2}\overrightarrow{d},-\frac{1}{2}\overrightarrow{b}) & w(-\frac{1}{2}\overrightarrow{d},-\frac{1}{2}\overrightarrow{c})\\
\end{array}%
\right)
\end{eqnarray}
form the semigroup of matrices satisfying the inequality
(\ref{eq.67}). This property can be used as criterion of the
separability. For example we take the two matrices
$M_1(\overrightarrow{a_1}\overrightarrow{b_1}\overrightarrow{c_1}\overrightarrow{d_1})$
and
$M_2(\overrightarrow{a_2}\overrightarrow{b_2}\overrightarrow{c_2}\overrightarrow{d_2})$.
We check that for both matrices the product
$F=M_1M_2(\overrightarrow{a_1}\overrightarrow{b_1}\overrightarrow{c_1}\overrightarrow{d_1}\overrightarrow{a_2}\overrightarrow{b_2}\overrightarrow{c_2}\overrightarrow{d_2})$
satisfies the inequality (\ref{eq.67}) for arbitrary directions
$(\overrightarrow{a_k}\overrightarrow{b_k}\overrightarrow{c_k}\overrightarrow{d_k})\quad
(k=1,2)$. This property can be generalized to any number of
directions $k=1,2...$. It is worthy  to note that the product of
two density matrices of two separable quantum states is not
density matrix of quantum state.
\section{Example of entangled states}
Let us take known example of entangled state of two qubits
\begin{eqnarray}\label{eq.78}
\rho=\frac{1}{2}\left(%
\begin{array}{cccc}
  1 & 0 & 0 & 1 \\
  0 & 0 & 0 & 0 \\
  0 & 0 & 0 & 0 \\
  1 & 0 & 0 & 1 \\
\end{array}%
\right).
\end{eqnarray}
We construct the tomogram of this state using (\ref{eq.55}),
(\ref{eq.56}). The result reads

\begin{eqnarray}
w(+\frac{1}{2},+\frac{1}{2},\overrightarrow{n_1},\overrightarrow{n_2})=\frac{1}{2}(\cos^2{\frac{\Theta_1}{2}}\cos^2{\frac{\Theta_2}{2}}+\sin^2{\frac{\Theta_1}{2}}\sin^2{\frac{\Theta_2}{2}})+\frac{1}{4}\sin\Theta_1\sin\Theta_2\cos{(\varphi_1+\varphi_2)};\nonumber\\
w(+\frac{1}{2},-\frac{1}{2},\overrightarrow{n_1},\overrightarrow{n_2})=\frac{1}{2}(\cos^2{\frac{\Theta_1}{2}}\sin^2{\frac{\Theta_2}{2}}+\sin^2{\frac{\Theta_1}{2}}\cos^2{\frac{\Theta_2}{2}})-\frac{1}{4}\sin\Theta_1\sin\Theta_2\cos{(\varphi_1+\varphi_2)};\nonumber\\
w(-\frac{1}{2},+\frac{1}{2},\overrightarrow{n_1},\overrightarrow{n_2})=\frac{1}{2}(\cos^2{\frac{\Theta_1}{2}}\sin^2{\frac{\Theta_2}{2}}+\sin^2{\frac{\Theta_1}{2}}\cos^2{\frac{\Theta_2}{2}})-\frac{1}{4}\sin\Theta_1\sin\Theta_2\cos{(\varphi_1+\varphi_2)};\nonumber\\
w(-\frac{1}{2},-\frac{1}{2},\overrightarrow{n_1},\overrightarrow{n_2})=\frac{1}{2}(\cos^2{\frac{\Theta_1}{2}}\cos^2{\frac{\Theta_2}{2}}+\sin^2{\frac{\Theta_1}{2}}\sin^2{\frac{\Theta_2}{2}})+\frac{1}{4}\sin\Theta_1\sin\Theta_2\cos{(\varphi_1+\varphi_2)}.\label{eq.79}
\end{eqnarray}
The matrix
$M(\overrightarrow{a},\overrightarrow{b},\overrightarrow{c},\overrightarrow{d})$
associated with the tomogram (\ref{eq.79}) has the 16 matrix
elements
\begin{eqnarray}
M_{11}=\frac{1}{2}(\cos^2{\frac{\Theta_a}{2}}\cos^2{\frac{\Theta_b}{2}}+\sin^2{\frac{\Theta_a}{2}}\sin^2{\frac{\Theta_b}{2}})+\frac{1}{4}\sin\Theta_a\sin\Theta_b\cos{(\varphi_a+\varphi_b)};\nonumber\\
M_{21}=\frac{1}{2}(\cos^2{\frac{\Theta_a}{2}}\sin^2{\frac{\Theta_b}{2}}+\sin^2{\frac{\Theta_a}{2}}\cos^2{\frac{\Theta_b}{2}})-\frac{1}{4}\sin\Theta_a\sin\Theta_b\cos{(\varphi_a+\varphi_b)};\nonumber\\
M_{31}=\frac{1}{2}(\cos^2{\frac{\Theta_a}{2}}\sin^2{\frac{\Theta_b}{2}}+\sin^2{\frac{\Theta_a}{2}}\cos^2{\frac{\Theta_b}{2}})-\frac{1}{4}\sin\Theta_a\sin\Theta_b\cos{(\varphi_a+\varphi_b)};\nonumber\\
M_{41}=\frac{1}{2}(\cos^2{\frac{\Theta_a}{2}}\cos^2{\frac{\Theta_b}{2}}+\sin^2{\frac{\Theta_a}{2}}\sin^2{\frac{\Theta_b}{2}})+\frac{1}{4}\sin\Theta_a\sin\Theta_b\cos{(\varphi_a+\varphi_b)};\nonumber\\
M_{12}=\frac{1}{2}(\cos^2{\frac{\Theta_a}{2}}\cos^2{\frac{\Theta_c}{2}}+\sin^2{\frac{\Theta_a}{2}}\sin^2{\frac{\Theta_c}{2}})+\frac{1}{4}\sin\Theta_a\sin\Theta_c\cos{(\varphi_a+\varphi_c)};\nonumber\\
M_{22}=\frac{1}{2}(\cos^2{\frac{\Theta_a}{2}}\sin^2{\frac{\Theta_c}{2}}+\sin^2{\frac{\Theta_a}{2}}\cos^2{\frac{\Theta_c}{2}})-\frac{1}{4}\sin\Theta_a\sin\Theta_c\cos{(\varphi_a+\varphi_c)};\nonumber\\
M_{32}=\frac{1}{2}(\cos^2{\frac{\Theta_a}{2}}\sin^2{\frac{\Theta_c}{2}}+\sin^2{\frac{\Theta_a}{2}}\cos^2{\frac{\Theta_c}{2}})-\frac{1}{4}\sin\Theta_a\sin\Theta_c\cos{(\varphi_a+\varphi_c)};\nonumber\\
M_{42}=\frac{1}{2}(\cos^2{\frac{\Theta_a}{2}}\cos^2{\frac{\Theta_c}{2}}+\sin^2{\frac{\Theta_a}{2}}\sin^2{\frac{\Theta_c}{2}})+\frac{1}{4}\sin\Theta_a\sin\Theta_c\cos{(\varphi_a+\varphi_c)};\nonumber\\
M_{13}=\frac{1}{2}(\cos^2{\frac{\Theta_d}{2}}\cos^2{\frac{\Theta_b}{2}}+\sin^2{\frac{\Theta_d}{2}}\sin^2{\frac{\Theta_b}{2}})+\frac{1}{4}\sin\Theta_d\sin\Theta_b\cos{(\varphi_d+\varphi_b)};\nonumber\\
M_{23}=\frac{1}{2}(\cos^2{\frac{\Theta_d}{2}}\sin^2{\frac{\Theta_b}{2}}+\sin^2{\frac{\Theta_d}{2}}\cos^2{\frac{\Theta_b}{2}})-\frac{1}{4}\sin\Theta_d\sin\Theta_b\cos{(\varphi_d+\varphi_b)};\nonumber\\
M_{33}=\frac{1}{2}(\cos^2{\frac{\Theta_d}{2}}\sin^2{\frac{\Theta_b}{2}}+\sin^2{\frac{\Theta_d}{2}}\cos^2{\frac{\Theta_b}{2}})-\frac{1}{4}\sin\Theta_d\sin\Theta_b\cos{(\varphi_d+\varphi_b)};\nonumber\\
M_{43}=\frac{1}{2}(\cos^2{\frac{\Theta_d}{2}}\cos^2{\frac{\Theta_b}{2}}+\sin^2{\frac{\Theta_d}{2}}\sin^2{\frac{\Theta_b}{2}})+\frac{1}{4}\sin\Theta_d\sin\Theta_b\cos{(\varphi_d+\varphi_b)};\nonumber\\
M_{14}=\frac{1}{2}(\cos^2{\frac{\Theta_d}{2}}\cos^2{\frac{\Theta_c}{2}}+\sin^2{\frac{\Theta_d}{2}}\sin^2{\frac{\Theta_c}{2}})+\frac{1}{4}\sin\Theta_d\sin\Theta_c\cos{(\varphi_d+\varphi_c)};\nonumber\\
M_{24}=\frac{1}{2}(\cos^2{\frac{\Theta_d}{2}}\sin^2{\frac{\Theta_c}{2}}+\sin^2{\frac{\Theta_d}{2}}\cos^2{\frac{\Theta_c}{2}})-\frac{1}{4}\sin\Theta_d\sin\Theta_c\cos{(\varphi_d+\varphi_c)};\nonumber\\
M_{34}=\frac{1}{2}(\cos^2{\frac{\Theta_d}{2}}\sin^2{\frac{\Theta_c}{2}}+\sin^2{\frac{\Theta_d}{2}}\cos^2{\frac{\Theta_c}{2}})-\frac{1}{4}\sin\Theta_d\sin\Theta_c\cos{(\varphi_d+\varphi_c)};\nonumber\\
M_{44}=\frac{1}{2}(\cos^2{\frac{\Theta_d}{2}}\cos^2{\frac{\Theta_c}{2}}+\sin^2{\frac{\Theta_d}{2}}\sin^2{\frac{\Theta_c}{2}})+\frac{1}{4}\sin\Theta_d\sin\Theta_c\cos{(\varphi_d+\varphi_c)}.\label{eq.80}
\end{eqnarray}
One can see that matrix $M$ (\ref{eq.80}) violates the condition
(\ref{eq.67}) which is Bell inequality for some angles and takes
maximal value $2\sqrt{2}$ which is Cirelson bound~\cite{II4}. It is
due to entanglement of the state (\ref{eq.78}). Violation of Bell
inequalities signals that the state is entangled. The product $M$ of
two matrices (\ref{eq.80}) corresponding to angles
$\Theta_a,\Theta_b,\Theta_c,\Theta_d,\varphi_a,\varphi_b,\varphi_c,\varphi_d$
for the first matrix $M_1$ and
$\Theta_{a^{'}},\Theta_{b^{'}},\Theta_{c^{'}},\Theta_{d^{'}},\varphi_{a^{'}},\varphi_{b^{'}},\varphi_{c^{'}},\varphi_{d^{'}}$
for the second matrix $M_2$, i.e. $M=M_1M_2$ must satisfy Bell
inequality (\ref{eq.67}) for separable state. These matrices form
semigroup which is sub-semigroup of all the stochastic matrices
constructed of means of tomograms of all the quantum states.

\section{Reduction of the qubit-qutrit separability property to Bell inequalities for two qubits.}
Here we demonstrate the new necessary condition of separability of
qubit-qutrit state using the probability representation of quantum
states. The idea of the construction is to find the qubit portrait
of the qutrit state discussed in previous sections. If one has the
probability distribution vector with three nonnegative components
\begin{eqnarray}\label{eq.s81}
\overrightarrow{W}=\left(%
\begin{array}{c}
  W_1 \\
  W_2 \\
  W_3 \\
\end{array}%
\right)
\end{eqnarray}
where $W_1+W_2+W_3=1$ the new probability distribution vector
$\overrightarrow{\rho}$ can be constructed
\begin{eqnarray}\label{eq.s82}
\overrightarrow{\rho}=\left(%
\begin{array}{c}
  \rho_1 \\
  \rho_2 \\
\end{array}%
\right)=\left(%
\begin{array}{c}
  W_1 \\
  W_2+W_3 \\
\end{array}%
\right)
\end{eqnarray}
It means that each three-dimensional distribution induces
two-dimensional ones. One can use all vectors.
\begin{eqnarray}\label{eq.s83}
\overrightarrow{\rho^{'}}=\left(%
\begin{array}{c}
  \rho_1^{'} \\
  \rho_2^{'} \\
\end{array}%
\right)=\left(%
\begin{array}{c}
  W_1+W_2 \\
  W_3 \\
\end{array}%
\right)
\end{eqnarray}
and
\begin{eqnarray}\label{eq.s84}
\overrightarrow{\rho^{''}}=\left(%
\begin{array}{c}
  \rho_1^{''} \\
  \rho_2^{''} \\
\end{array}%
\right)=\left(%
\begin{array}{c}
  W_1+W_3 \\
  W_2 \\
\end{array}%
\right)
\end{eqnarray}
Let us consider simply separable state of qubit-qutrit system with
density operator $\hat{\rho}(1,2)=\hat{\rho}(1)\bigotimes
\hat{\rho}(2)$. Then the tomogram of this state is the probability
distribution of the form
\begin{eqnarray}\label{eq.s85}
w(m_1,\overrightarrow{n_1},m_2,\overrightarrow{n_2})=w_1(m_1,\overrightarrow{n_1})W(m_2,\overrightarrow{n_2}).
\end{eqnarray}
Here the spin projections $m_1$ take values $-1/2$, $+1/2$ and
spin projections $m_2$ take values $-1$, $+1$, $0$. In the form of
6-dimensional vector the tomogram (\ref{eq.s85}) can be rewritten
as
\begin{eqnarray}\label{eq.s86}
\overrightarrow{w}(\overrightarrow{n_1},\overrightarrow{n_2})=\overrightarrow{w_{\frac{1}{2}}}(\overrightarrow{n_1})\bigotimes
\overrightarrow{W_1}(\overrightarrow{n_2}),
\end{eqnarray}
where
\begin{eqnarray}\label{eq.s87}
\overrightarrow{w_{\frac{1}{2}}}=\left(%
\begin{array}{c}
  w_1(\overrightarrow{n_1}) \\
  w_2(\overrightarrow{n_1}) \\
\end{array}%
\right),
\end{eqnarray}
and
\begin{eqnarray}\label{eq.s88}
\overrightarrow{W_1}(\overrightarrow{n_2})=\left(%
\begin{array}{c}
  W_1(\overrightarrow{n_2}) \\
  W_2(\overrightarrow{n_2})  \\
  W_3(\overrightarrow{n_2})  \\
\end{array}%
\right).
\end{eqnarray}
Thus one has
\begin{eqnarray}\label{eq.s89}
\overrightarrow{w}(\overrightarrow{n_1},\overrightarrow{n_2})=\left(%
\begin{array}{c}
  w_1(\overrightarrow{n_1})W_1(\overrightarrow{n_2}) \\
  w_1(\overrightarrow{n_1})W_2(\overrightarrow{n_2}) \\
  w_1(\overrightarrow{n_1})W_3(\overrightarrow{n_2}) \\
  w_2(\overrightarrow{n_1})W_1(\overrightarrow{n_2}) \\
  w_2(\overrightarrow{n_1})W_2(\overrightarrow{n_2}) \\
  w_2(\overrightarrow{n_1})W_3(\overrightarrow{n_2}) \\\end{array}%
\right).
\end{eqnarray}
Now we apply the described ansatz of reduction of three
dimensional distributions to two dimensional ones. We get frow $m$
(\ref{eq.s88}) the vector
\begin{eqnarray}\label{eq.s90}
\overrightarrow{\rho_1}(\overrightarrow{n_2})=\left(%
\begin{array}{c}
  W_1(\overrightarrow{n_2}) \\
  W_2(\overrightarrow{n_2})+W_3(\overrightarrow{n_2}) \\
\end{array}%
\right).
\end{eqnarray}
This reduction induces the reduction of the 6-vector
(\ref{eq.s89}) to the 4-vector
\begin{eqnarray}\label{eq.s91}
\overrightarrow{\rho}(\overrightarrow{n_1},\overrightarrow{n_2})=\left(%
\begin{array}{c}
  w_1(\overrightarrow{n_1})W_1(\overrightarrow{n_2}) \\
  w_1(\overrightarrow{n_1})(W_2(\overrightarrow{n_2})+W_3(\overrightarrow{n_2})) \\
  w_2(\overrightarrow{n_1})W_1(\overrightarrow{n_2}) \\
  w_2(\overrightarrow{n_1})(W_2(\overrightarrow{n_2})+W_3(\overrightarrow{n_2})) \\
\end{array}%
\right).
\end{eqnarray}
One has the simple observation. If the tomogram is simply
separable the reduced distribution vector
$\overrightarrow{\rho}(\overrightarrow{n_1},\overrightarrow{n_2})$
is also simply separable distribution. From this property if
follows the same property for a convex sum of simply separable
distributions. One has for separable quantum state of qubit-qutrit
system the following property of its spin tomogram. Let this spin
tomogram be given by a probability distribution
$w(m_1,\overrightarrow{n_1},m_2,\overrightarrow{n_2})$ which
corresponds either to separable or entangled state. Let us denote
this tomogram by the vector
\begin{eqnarray}\label{eq.s92}
\overrightarrow{w}(\overrightarrow{n_1},\overrightarrow{n_2})=\left(%
\begin{array}{c}
  w(+\frac{1}{2},\overrightarrow{n_1},+1,\overrightarrow{n_2}) \\
  w(+\frac{1}{2},\overrightarrow{n_1},0,\overrightarrow{n_2}) \\
  w(+\frac{1}{2},\overrightarrow{n_1},-1,\overrightarrow{n_2}) \\
  w(-\frac{1}{2},\overrightarrow{n_1},+1,\overrightarrow{n_2}) \\
  w(-\frac{1}{2},\overrightarrow{n_1},0,\overrightarrow{n_2}) \\
  w(-\frac{1}{2},\overrightarrow{n_1},-1,\overrightarrow{n_2}) \\
\end{array}%
\right).
\end{eqnarray}
Then we introduce the 4-vector
\begin{eqnarray}\label{eq.s93}
\overrightarrow{\rho}(\overrightarrow{n_1},\overrightarrow{n_2})=\left(%
\begin{array}{c}
  w(+\frac{1}{2},\overrightarrow{n_1},+1,\overrightarrow{n_2}) \\
  w(+\frac{1}{2},\overrightarrow{n_1},0,\overrightarrow{n_2})+w(+\frac{1}{2},\overrightarrow{n_1},-1,\overrightarrow{n_2}) \\
  w(-\frac{1}{2},\overrightarrow{n_1},+1,\overrightarrow{n_2}) \\
  w(-\frac{1}{2},\overrightarrow{n_1},0,\overrightarrow{n_2})+w(-\frac{1}{2},\overrightarrow{n_1},-1,\overrightarrow{n_2})\\
\end{array}%
\right).
\end{eqnarray}
Now we apply the criterion of separability used for two qubit
states discussed in previous sections. It means that we construct
stochastic 4x4-matrix where in the column one has the components
of the vectors (\ref{eq.s93}) with corresponding vectors
$\overrightarrow{n_1},\overrightarrow{n_2}$
\begin{eqnarray}\label{eq.s94}
P(\overrightarrow{a},\overrightarrow{b},\overrightarrow{c},\overrightarrow{d})=\parallel
\overrightarrow{\rho}(\overrightarrow{a},\overrightarrow{b})\overrightarrow{\rho}(\overrightarrow{a},\overrightarrow{c})\overrightarrow{\rho}(\overrightarrow{d},\overrightarrow{b})\overrightarrow{\rho}(\overrightarrow{d},\overrightarrow{c})\parallel.
\end{eqnarray}
We get the result. If the matrix elements of the matrix
(\ref{eq.s94}) violate the Bell inequality the qubit-qutrit state
is entangled. The fulfilling of the Bell inequality (\ref{eq.76})
is necessary condition of the separability of the qubit-qutrit
state.

\section{Qubit-qutrit and two qutrits}
We present here two examples of entangled states. Let density
matrix of qubit-qutrit state in standard basis $|1/2,m_1>|1,m_2>$
have the form
\begin{eqnarray}\label{eq.G1}
\rho=\frac{1}{2}\left(%
\begin{array}{cccccc}
  1 & 0 & 0 & 0 & 0 & 1 \\
  0 & 0 & 0 & 0 & 0 & 0 \\
  0 & 0 & 0 & 0 & 0 & 0 \\
  0 & 0 & 0 & 0 & 0 & 0 \\
  0 & 0 & 0 & 0 & 0 & 0 \\
  1 & 0 & 0 & 0 & 0 & 1 \\
\end{array}%
\right).
\end{eqnarray}
Two unitary matrices transforming qubits
\begin{eqnarray}
U_{11}=e^{\frac{\iota \varphi_1}{2}}\cos{\frac{\theta_1}{2}};\nonumber\\
U_{12}=\iota e^{\frac{\iota \varphi_1}{2}}\sin{\frac{\theta_1}{2}};\nonumber\\
U_{21}=\iota e^{\frac{-\iota \varphi_1}{2}}\sin{\frac{\theta_1}{2}};\nonumber\\
U_{22}=e^{\frac{-\iota\varphi_1}{2}}\cos{\frac{\theta_1}{2}};\label{eq.G2}
\end{eqnarray}
and qutrits
\begin{eqnarray}
V_{11}=e^{\iota \varphi_2}\cos^2{\frac{\theta_2}{2}};\nonumber\\
V_{12}=\iota e^{\iota \varphi_2}\frac{\sin{\Theta_2}}{\sqrt{2}};\nonumber\\
V_{13}=-e^{\iota \varphi_2}\sin^2{\frac{\theta_2}{2}};\nonumber\\
V_{21}=\iota \frac{\sin{\Theta_2}}{\sqrt{2}};\nonumber\\
V_{22}=\cos{\Theta_2};\nonumber\\
V_{23}=\iota \frac{\sin{\Theta_2}}{\sqrt{2}};\nonumber\\
V_{31}=-e^{-\iota \varphi_2}\sin^2{\frac{\theta_2}{2}};\nonumber\\
V_{32}=\iota e^{-\iota \varphi_2}\frac{\sin{\Theta_2}}{\sqrt{2}};\nonumber\\
V_{33}=e^{-\iota
\varphi_2}\cos^2{\frac{\theta_2}{2}};\label{eq.G3}
\end{eqnarray}
can be used to construct the 6x6-matrices $U\bigotimes V$ and
$U^{\dag}\bigotimes V^{\dag}$. The diagonal matrix elements of the
matrix
\begin{eqnarray}\label{eq.G4}
[(U^{\dag}\bigotimes V^{\dag})\rho(U\bigotimes
V)]_{m_1m_2,m_1m_2}=w(m_1,\overrightarrow{n_1},m_2,\overrightarrow{n_2})
\end{eqnarray}
provide the spin tomogram of the state (\ref{eq.G1}). Here the two
vectors are determined by angles
$\theta_1,\varphi_1,\theta_2,\varphi_2$ as
$\overrightarrow{n_1}=(\sin{\Theta_1}\cos{\varphi_1},\sin{\Theta_1}\sin{\varphi_1},\cos{\Theta_1})$,
$\overrightarrow{n_2}=(\sin{\Theta_2}\cos{\varphi_2},\sin{\Theta_2}\sin{\varphi_2},\cos{\Theta_2})$.\\
One has
\begin{eqnarray}
w(+\frac{1}{2},\overrightarrow{n_1},+1,\overrightarrow{n_2})=\frac{1}{2}|U_{11}V_{11}+U_{21}V_{31}|^2;\nonumber\\
w(+\frac{1}{2},\overrightarrow{n_1},0,\overrightarrow{n_2})=\frac{1}{2}|U_{11}V_{12}+U_{21}V_{32}|^2; \nonumber\\
w(+\frac{1}{2},\overrightarrow{n_1},-1,\overrightarrow{n_2})=\frac{1}{2}|U_{11}V_{13}+U_{21}V_{33}|^2; \nonumber\\
w(-\frac{1}{2},\overrightarrow{n_1},+1,\overrightarrow{n_2})=\frac{1}{2}|U_{12}V_{11}+U_{22}V_{31}|^2; \nonumber\\
w(-\frac{1}{2},\overrightarrow{n_1},0,\overrightarrow{n_2})=\frac{1}{2}|U_{12}V_{12}+U_{22}V_{32}|^2; \nonumber\\
w(-\frac{1}{2},\overrightarrow{n_1},-1,\overrightarrow{n_2})=\frac{1}{2}|U_{12}V_{13}+U_{22}V_{33}|^2.\label{eq.G5}
\end{eqnarray}
Applying the reduction ansatz we get the 4x4-matrix
(\ref{eq.s94}). Calculating the modulus of trace of product of
this matrix and the matrix $I$ given by  (\ref{eq.63}) we get the
expression which we denote as
\begin{eqnarray}\label{eq.G6}
B=|\sin\Theta_a(\sin^2\Theta_b\sin\Phi_{ab}+\sin^2\Theta_c\sin\Phi_{ac})+\sin\Theta_d(\sin^2\Theta_b\sin\Phi_{db}-\sin^2\Theta_c\sin\Phi_{dc})|.
\end{eqnarray}
Here $\Phi_{ab}=\varphi_a+2\varphi_b$,\quad
$\Phi_{ac}=\varphi_a+2\varphi_c$,\quad
$\Phi_{db}=\varphi_d+2\varphi_b$,\quad
$\Phi_{dc}=\varphi_d+2\varphi_c$. One can check that for
parameters
\begin{eqnarray}
\Theta_a=\frac{\pi}{2},\quad \Theta_b=\frac{\pi}{2}, \quad \Theta_c=\frac{\pi}{2}, \quad \Theta_d=\frac{\pi}{2}, \nonumber\\
\Phi_{ab}=\frac{\pi}{2},\quad \Phi_{dc}=-\frac{\pi}{4},\quad
\Phi_{ac}=\frac{\pi}{4},\quad \Phi_{db}=0 \label{eq.G7}
\end{eqnarray}
the value $B$ (\ref{eq.G6}) is larger than 2, namely
\begin{eqnarray}\label{eq.G8}
B=1+\sqrt{2}.
\end{eqnarray}
If means that the qubit-qutrit state is entangled. We know this
fact because the density matrix (\ref{eq.G1}) corresponds to pure
entangled state
$|\Psi\rangle=\frac{1}{\sqrt{2}}(|+\frac{1}{2}\rangle|+1\rangle+|-\frac{1}{2}\rangle|-1\rangle)$.
For two qutrit entangled state with 9x9-density matrix with 72
matrix elements equal to zero except 9 matrix elements
\begin{eqnarray}\label{eq.G9}
\rho_{11}=\rho_{15}=\rho_{19}=\rho_{51}=\rho_{55}=\rho_{59}=\rho_{91}=\rho_{95}=\rho_{99}=\frac{1}{3}
\end{eqnarray}
the spin tomogram can be calculated by the same method using two
3x3-matrices $U$ and $V$ given by the same relations
(\ref{eq.G3}). But the matrix elements of the matrix $U$ are taken
to depend on angles $\varphi_1$ and $\Theta_1$. We get the vector
$\overrightarrow{w}(\overrightarrow{n_1},\overrightarrow{n_2})$
with nine components:
\begin{eqnarray}
w(+1,\overrightarrow{n_1},+1,\overrightarrow{n_2})=\frac{1}{3}|\sum_{j=1}^{3}U_{j1}V_{j1}|^2;\nonumber\\
w(+1,\overrightarrow{n_1},0,\overrightarrow{n_2})=\frac{1}{3}|\sum_{j=1}^{3}U_{j1}V_{j2}|^2;\nonumber\\
w(+1,\overrightarrow{n_1},-1,\overrightarrow{n_2})=\frac{1}{3}|\sum_{j=1}^{3}U_{j1}V_{j3}|^2;\nonumber\\
w(0,\overrightarrow{n_1},+1,\overrightarrow{n_2})=\frac{1}{3}|\sum_{j=1}^{3}U_{j2}V_{j1}|^2;\nonumber\\
w(0,\overrightarrow{n_1},0,\overrightarrow{n_2})=\frac{1}{3}|\sum_{j=1}^{3}U_{j2}V_{j2}|^2;\nonumber\\
w(0,\overrightarrow{n_1},-1,\overrightarrow{n_2})=\frac{1}{3}|\sum_{j=1}^{3}U_{j2}V_{j3}|^2;\nonumber\\
w(-1,\overrightarrow{n_1},+1,\overrightarrow{n_2})=\frac{1}{3}|\sum_{j=1}^{3}U_{j3}V_{j1}|^2;\nonumber\\
w(-1,\overrightarrow{n_1},0,\overrightarrow{n_2})=\frac{1}{3}|\sum_{j=1}^{3}U_{j3}V_{j2}|^2;\nonumber\\
w(-1,\overrightarrow{n_1},-1,\overrightarrow{n_2})=\frac{1}{3}|\sum_{j=1}^{3}U_{j3}V_{j3}|^2.\label{eq.G10}
\end{eqnarray}
We construct the qubit portrait of this state. One of 4-vectors
$\overrightarrow{P}(\overrightarrow{n_1},\overrightarrow{n_2})$ of
this portrait has the components
\begin{eqnarray}
P_1(\overrightarrow{n_1},\overrightarrow{n_2})=w(+1,\overrightarrow{n_1},+1,\overrightarrow{n_2})\nonumber\\
P_2(\overrightarrow{n_1},\overrightarrow{n_2})=w(+1,\overrightarrow{n_1},0,\overrightarrow{n_2})+w(+1,\overrightarrow{n_1},-1,\overrightarrow{n_2})\nonumber\\
P_3(\overrightarrow{n_1},\overrightarrow{n_2})=w(0,\overrightarrow{n_1},+1,\overrightarrow{n_2})+w(-1,\overrightarrow{n_1},+1,\overrightarrow{n_2})\nonumber\\
P_4(\overrightarrow{n_1},\overrightarrow{n_2})=w(0,\overrightarrow{n_1},0,\overrightarrow{n_2})+w(0,\overrightarrow{n_1},-1,\overrightarrow{n_2})+w(-1,\overrightarrow{n_1},0,\overrightarrow{n_2})+w(-1,\overrightarrow{n_1},-1,\overrightarrow{n_2})\label{eq.G11}
\end{eqnarray}
Using (\ref{eq.G10}) and (\ref{eq.G11}) and taking pairs
$(\overrightarrow{n_1})=\overrightarrow{a},(\overrightarrow{n_2})=\overrightarrow{b}$,
\quad
$(\overrightarrow{n_1})=\overrightarrow{a},(\overrightarrow{n_2})=\overrightarrow{c}$,
\quad
$(\overrightarrow{n_1})=\overrightarrow{d},(\overrightarrow{n_2})=\overrightarrow{b}$,
\quad
$(\overrightarrow{n_1})=\overrightarrow{d},(\overrightarrow{n_2})=\overrightarrow{c}$
one can construct the 4x4-matrix (\ref{eq.s94}). Calculating the
modulus of trace of product of matrix (\ref{eq.63}) with the
obtained matrix we get the value of $B$ of the form
\begin{eqnarray}
B=\frac{1}{2}|((\cos\Theta_b+1)^2-2)(\cos\Theta_a+\cos\Theta_d)+\nonumber\\
+((\cos\Theta_c+1)^2-2)(\cos\Theta_a-\cos\Theta_d)-\nonumber\\
-\sin^2\Theta_b(\sin\Phi_{ab}\sin\Theta_a+\sin\Phi_{db}\sin\Theta_d)-\nonumber\\
-\sin^2\Theta_c(\sin\Phi_{ac}\sin\Theta_a+\sin\Phi_{dc}\sin\Theta_d)|\label{eq.G12}
\end{eqnarray}
One can check that for angles
\begin{eqnarray}
\varphi_a=2\pi, \quad\varphi_b=-\frac{\pi}{8},\quad
\varphi_c=\frac{\pi}{8},\quad
\varphi_d=0,\nonumber\\
\Theta_a=0,\quad\Theta_b=\frac{\pi}{2},\quad\Theta_c=\frac{\pi}{2},\quad\Theta_d=\frac{\pi}{2}.\label{eq.G13}
\end{eqnarray}
the value of $B$ is $(1+\sqrt{2})>2$. It corresponds to entangled
two qutrit state.

\section{General reduction criterion of separability}
Now we use the experience with discussed qubit-qutrit system to
formulate a general criterion of separability for a state of
bipartite quantum system. The criterion is based on the property
of a separable state tomogram of a bipartite system. Let us take
for simplicity a two qudite separable state with the tomogram of
the form (\ref{eq.68}). Let us associate with this tomogram the
joint probability distribution given as four nonnegative numbers
\begin{eqnarray}
\tilde{w}(M_1=j_1,M_2=j_2,\overrightarrow{n_1},\overrightarrow{n_2})=w(j_1,j_2,\overrightarrow{n_1},\overrightarrow{n_2});\nonumber\\
\tilde{w}(M_1=j_1,M_2=j_2-1,\overrightarrow{n_1},\overrightarrow{n_2})=\sum_{m_2=-j_2}^{j_2-1}w(j_1,m_2,\overrightarrow{n_1},\overrightarrow{n_2});\nonumber\\
\tilde{w}(M_1=j_1-1,M_2=j_2,\overrightarrow{n_1},\overrightarrow{n_2})=\sum_{m_1=-j_1}^{j_1-1}w(m_1,j_2,\overrightarrow{n_1},\overrightarrow{n_2});\nonumber\\
\tilde{w}(M_1=j_1-1,M_2=j_2-1,\overrightarrow{n_1},\overrightarrow{n_2})=\sum_{m_1=-j_1}^{j_1-1}\sum_{m_2=-j_2}^{j_2-1}w(m_1,m_2,\overrightarrow{n_1},\overrightarrow{n_2}).\nonumber\\
\end{eqnarray}
Here $M_1$ takes two values $j_1$ and $j_1-1$ and $M_2$ takes the
values $j_2$ and $j_2-1$. We will reinterpret the obtained joint
probability distribution as a two-qubit "tomogram". Due to this
the Bell inequality is fulfilled for the probability distribution
if the initial two-qudit state is separable. We used ansatz of
obtaining the reduced joint probability distribution by summing
the probabilities in initial probability distribution with larger
number of possible events (or measurements). But the separability
of the initial quantum state is preserved in process of such
summing in the sense that if initial tomographic probability
distribution looks as a convex sum of products of two
distributions the reduced distribution is also the convex sum of
the product of two probability distributions. The obtained result
can be formulated as the following reduction criterion of
separability. The necessary condition of separability of bipartite
system state is the separability property of the reduced state
tomogram The fulfilling of Bell inequalities for reduced state
tomogram is necessary condition of separability of the quantum
state under study. One can give a recipe for studying the
separability of a given state of bipartite system. First step is
to obtain the tomogram of the state. Than one has to reduce this
tomogram by summing over all such events to get the "tomogram" of
two qubit. Then one checks the fulfilling the Bell inequality for
the obtained reduced tomogram. If it is violated the initial state
is entangled.

\section{Conclusion}
To conclude we summarize the main results of our work. We shown
that the qudit states can be mapped onto probability distributions
which are the points on the simplex. The probability distributions
can be considered as vectors. The stochastic and bistochastic
matrices can be constructed using these vectors as columns of the
matrices. Both stochastic and bistochastic matrices form
semigroups. The invertible map of probability distributions onto
bistochastic matrix was used to construct star-product of the
probability distributions. For qudit tomograms we introduced the
notion of qubit portrait. We shown that the necessary condition of
separability of bipartite qudit state is separability of its qubit
portrait . The Bell inequality violation for qubit portrait of
bipartite system state (both for qudit states and for continuous
variables) means that the system state is entangled. Examples of
entangled qubit-qutrit state and two-qutrit state were considered
using the method of constructing the qubit portrait of the states.
The method can be generalized for multiqudit systems.

\end{document}